\documentstyle[jkas35]{article}

\runningauthor{RYU AND KANG}
\runningtitle{CLUSTERS OF GALAXIES}
\beginpage{1}
\endpage{8}

\begin{document}

\title{CLUSTERS OF GALAXIES: SHOCK WAVES AND COSMIC RAYS}

\author{Dongsu Ryu}

\address{Department of Astronomy \& Space Science,
Chungnam National University, Daejeon 305-764, Korea\\
{\it E-mail: ryu@canopus.chungnam.ac.kr}}

\author{and}

\author{Hyesung kang}

\address{Department of Earth Sciences, Pusan National University,
Pusan 609-735, Korea\\
{\it E-mail: kang@uju.es.pusan.ac.kr}}

\author{~}

\address{\normalsize{\it (Received October, 31, 2002; Accepted ???. ??,2002)}}

\abstract{
Recent observations of galaxy clusters in radio and X-ray indicate that
cosmic rays and magnetic fields may be energetically important
in the intracluster medium. According to the estimates based on theses
observational studies, the combined pressure of these two components
of the intracluster medium may range 
between $10\% \sim 100 \%$ of gas pressure, 
although their total energy is probably time dependent.  
Hence, these non-thermal components may have influenced the formation and 
evolution of cosmic structures, and may provide unique and vital diagnostic 
information through various radiations emitted via their interactions with 
surrounding matter and cosmic background photons. 
We suggest that shock waves associated with cosmic structures, 
along with individual sources such as active galactic nuclei 
and radio galaxies, supply the cosmic rays and
magnetic fields to the intracluster medium
and to surrounding large scale structures. 
In order to study 1) the properties of cosmic shock waves emerging during
the large scale structure formation of the universe, 
and 2) the dynamical influence of cosmic rays, which were ejected by 
AGN-like sources into the intracluster medium,
on structure formation, 
we have performed two sets of N-body/hydrodynamic simulations of cosmic 
structure formation.
In this contribution, we report the preliminary results of these simulations.}

\maketitle

\section{INTRODUCTION}

Formation and evolution of cosmic structures including clusters of galaxies
have been driven by the gravity of matter, and the matter continues to
accrete onto the cosmic structures as a part of hierarchical structure
formation in the cold dark matter paradigm. The
accreting matter has a typical velocity up to $\sim {\rm a~few}~10^3
{\rm km~s}^{-1}$, and its baryonic component ends as accretions shocks of 
Mach number from a few to a few hundred (see, e.g., Ryu \& Kang 1997a). 
The gravitational energy of the gas is dissipated via collisionless shocks, 
heating the intracluster gas to $10^6-10^8$K.
Cosmic shocks, along with the hierarchical clustering of predominant dark
matter, generate complex flow patterns of gas inside cosmic structures,
which, in turn, induce weaker, internal shocks. 
As a result, shock waves formed during the large scale structure formation 
span a wide range of scales from a few kpc to a few Mpc, and have
a wide range of Mach numbers from one to a few hundred 
(see, e.g., Ryu \& Kang 1997b, Miniati et al. 2000). 
The shocks can heat up the gas to the temperature ranging from 
$\sim 10^4-10^5$ K in pancakes up to $\sim 10^8$ K in clusters 
(see, e.g., Kang et al.  1994, Cen \& Ostriker 1999).

In addition to heating gas, the shock waves are likely to accelerate high
energy particles (cosmic rays) through the so-called diffusive shock
acceleration, provided that there exist weak magnetic fields in
the intergalactic space (see, e.g., Blandford \& Eichler 1987, Kang, Ryu,
\& Jones 1996). 
Extended regions populated by cosmic ray 
electrons have been observed in some clusters for more than thirty
years through diffuse, nonthermal radio emissions (see e.g., Kim et al.
1989). Although cosmic ray protons produce $\gamma$-rays
via $\pi^0$ decay following inelastic collisions with gas nuclei, such
$\gamma$-rays have not yet been detected (Sreekumar et al. 1996).
The observations of nonthermal radiations from cosmic ray electrons in
clusters, however, suggest that cosmic ray protons could be
energetically important, and possibly in energy equipartition with gas.
Cosmic shock waves may be the main sources of such cosmic rays, while
active galactic nuclei including radio galaxies should also provide
a significant amount of cosmic rays to the intracluster medium.

Cosmic shock waves could serve also as sites for generation of weak seeds
of magnetic fields by the Biermann battery mechanism. It was proposed that
these seeds could be amplified to strong magnetic fields of up to a few
$\mu{\rm G}$ in clusters, if flows there can be described as the Kolmogoroff
turbulence (Kulsrud et al. 1997). Although further development into coherent
magnetic fields is unclear, since there is as yet no detailed theory capable
of describing this process, observations suggest the existence of cluster
magnetic fields of a few $\mu{\rm G}$ strength and $\sim 10$ kpc coherent
length (see, e.g., Clarke et al. 2001).

We have studied the properties of cosmic shock waves 
and the dynamical roles of cosmic rays in clusters by high-resolution
cosmological hydrodynamic simulations. For the latter, we adopted an AGN
(active galactic nucleus) origin model in which cosmic ray energy
is deposited into the intracluster medium by AGNs, 
while the cosmic rays accelerated at cosmic shocks are ignored. 
This is because there is yet no practical numerical scheme which can
follow the non-linear diffusive shock acceleration of cosmic rays in
multi-dimensional simulations. In the next section we describe 
a cosmological hydrodynamic simulation and discuss the statistical
properties and roles of shock waves in cosmic structures. 
The final section briefs simulations which include the dynamical influences
of cosmic rays in the course of formation and evolution of cosmic structures.

\section{SHOCKS FROM A HIGH-RESOLUTION COSMOLOGY SIMULATION}

\begin {figure*}
\vskip -2cm
\centerline{\epsfysize=19cm\epsfbox{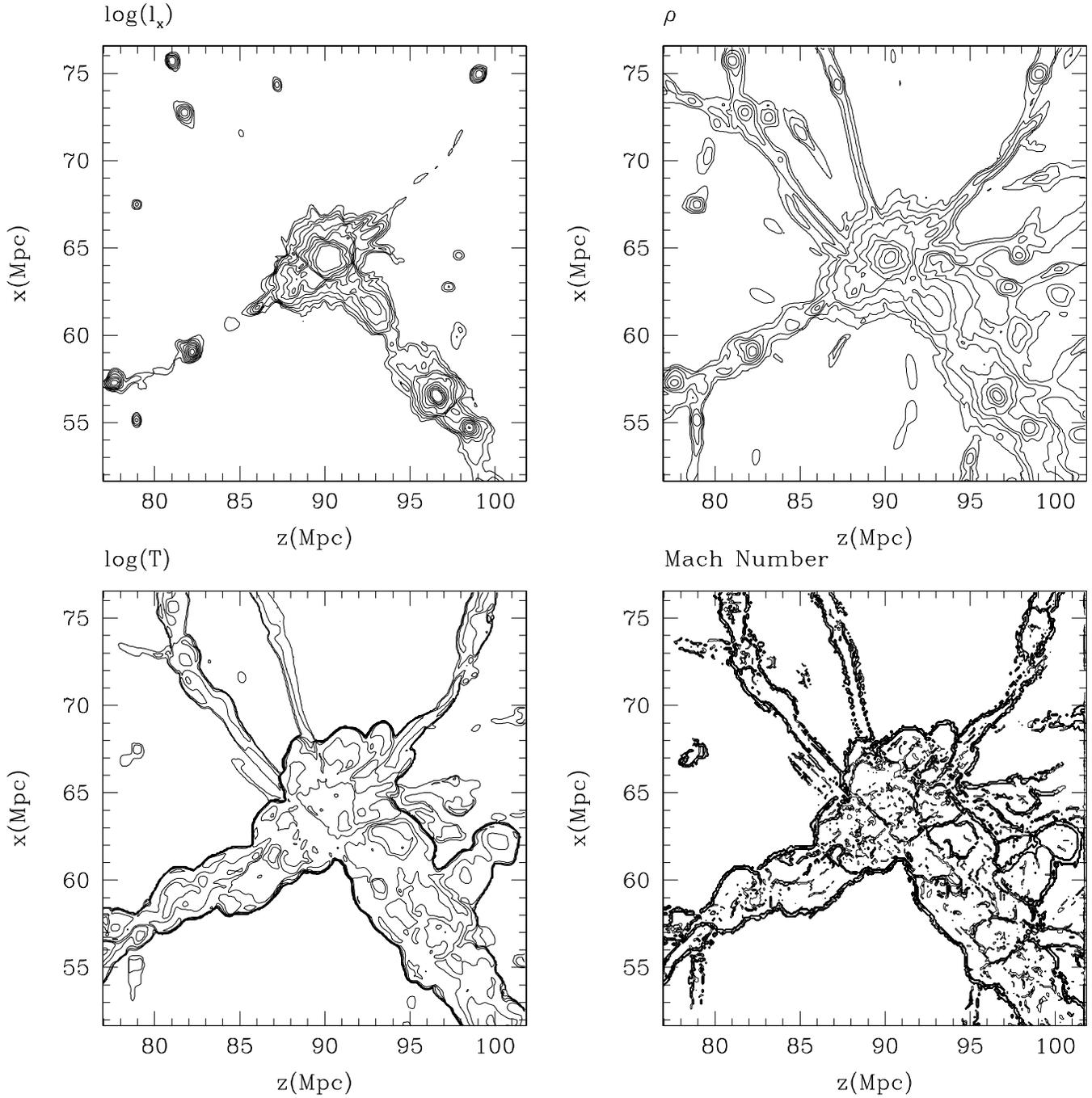}}
\vskip -0.5cm
\caption{Typical cosmic structures through a two-dimensional slice of
$(25 h^{-1} {\rm Mpc})^2$ out of the whole $(100 h^{-1} {\rm Mpc})^3$
box in the $\Lambda$CDM simulation with $1024^3$ grid cells.
Bremsstrahlung X-ray emissivity, gas density, gas temperature and
shock Mach number distributions are shown.}
\end{figure*}

\begin {figure*}
\vskip 0cm
\centerline{\epsfysize=7cm\epsfbox{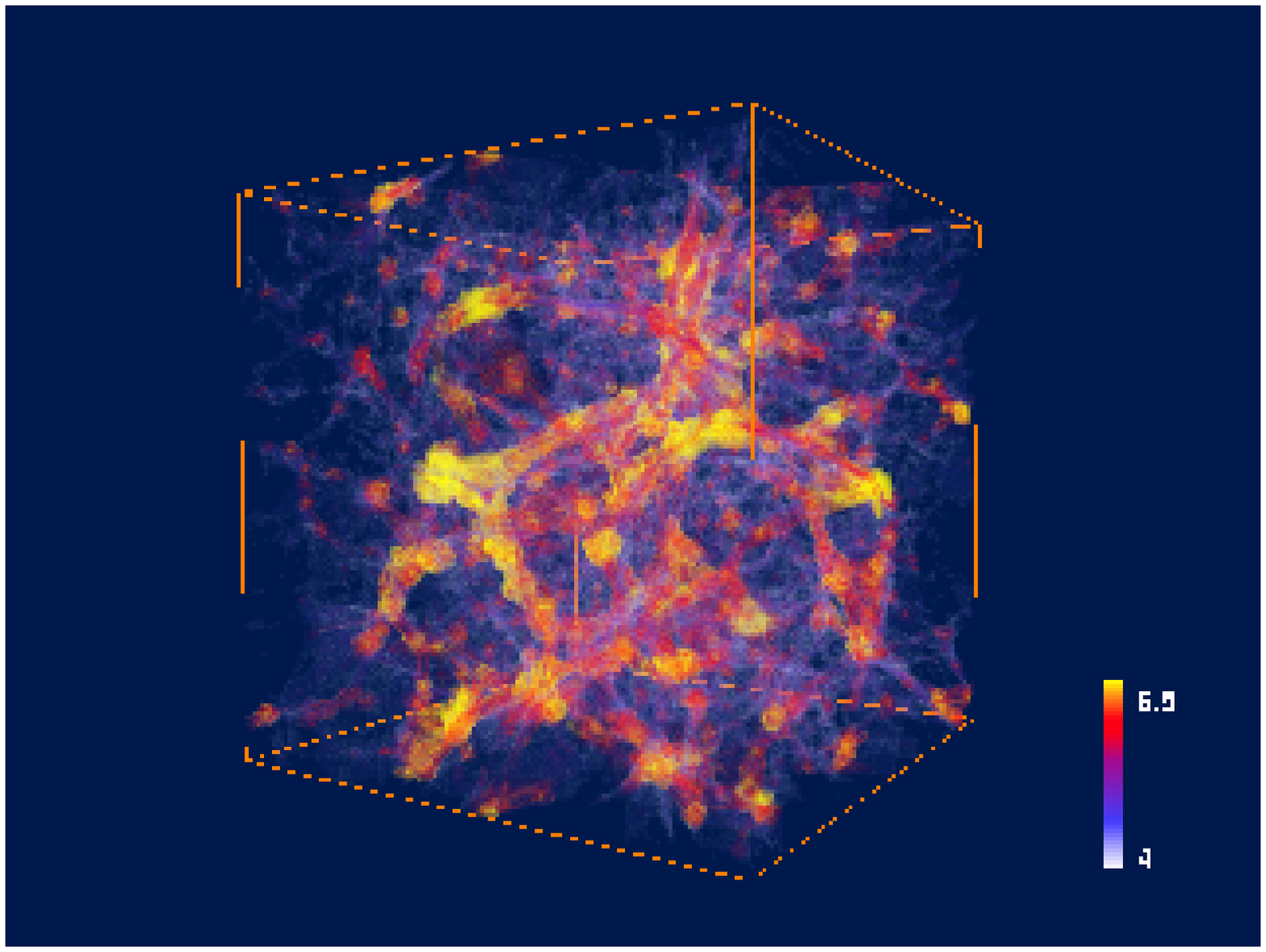}}
\vskip 0cm
\centerline{\epsfysize=7cm\epsfbox{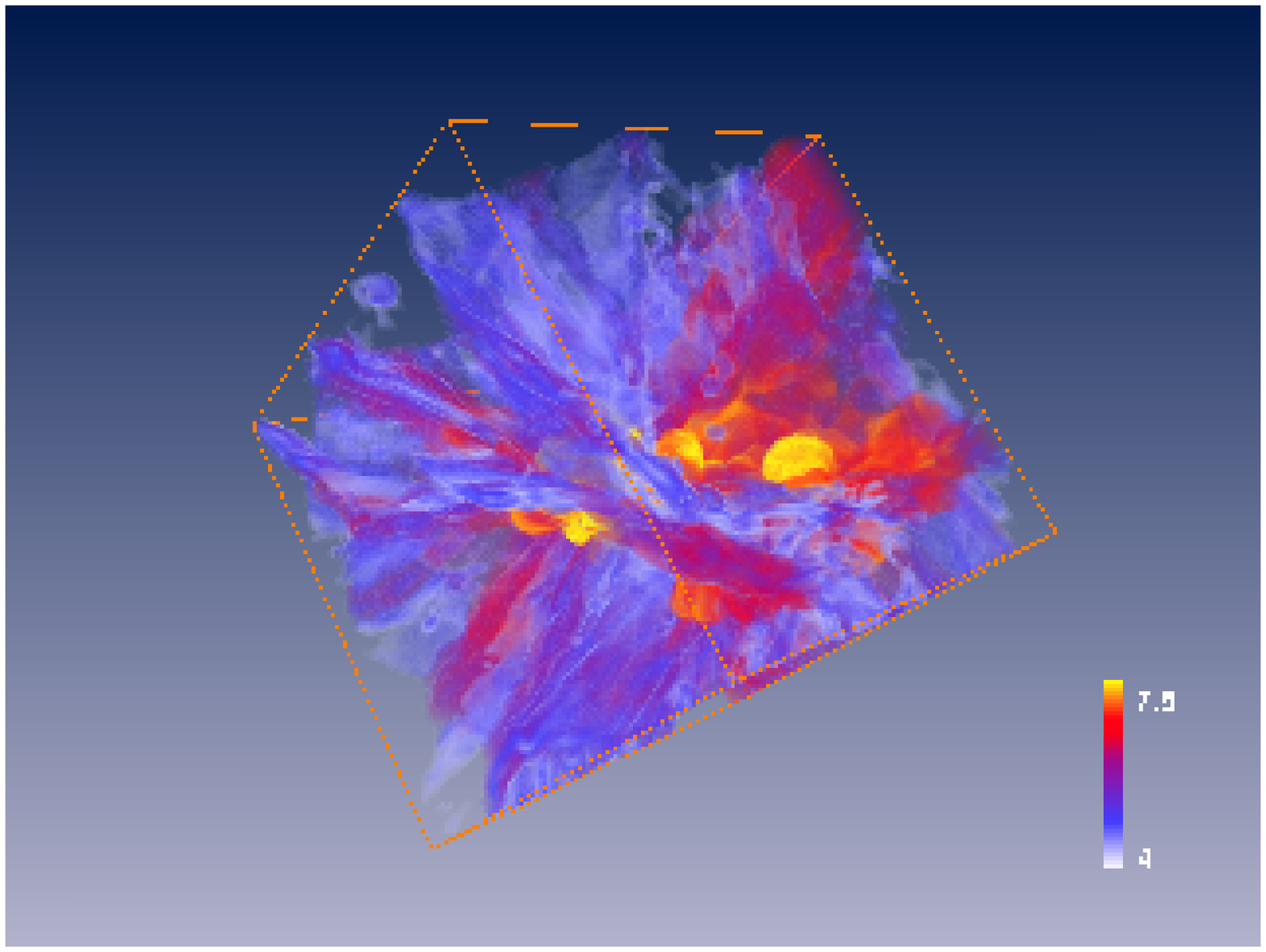}}
\vskip 0cm
\centerline{\epsfysize=7cm\epsfbox{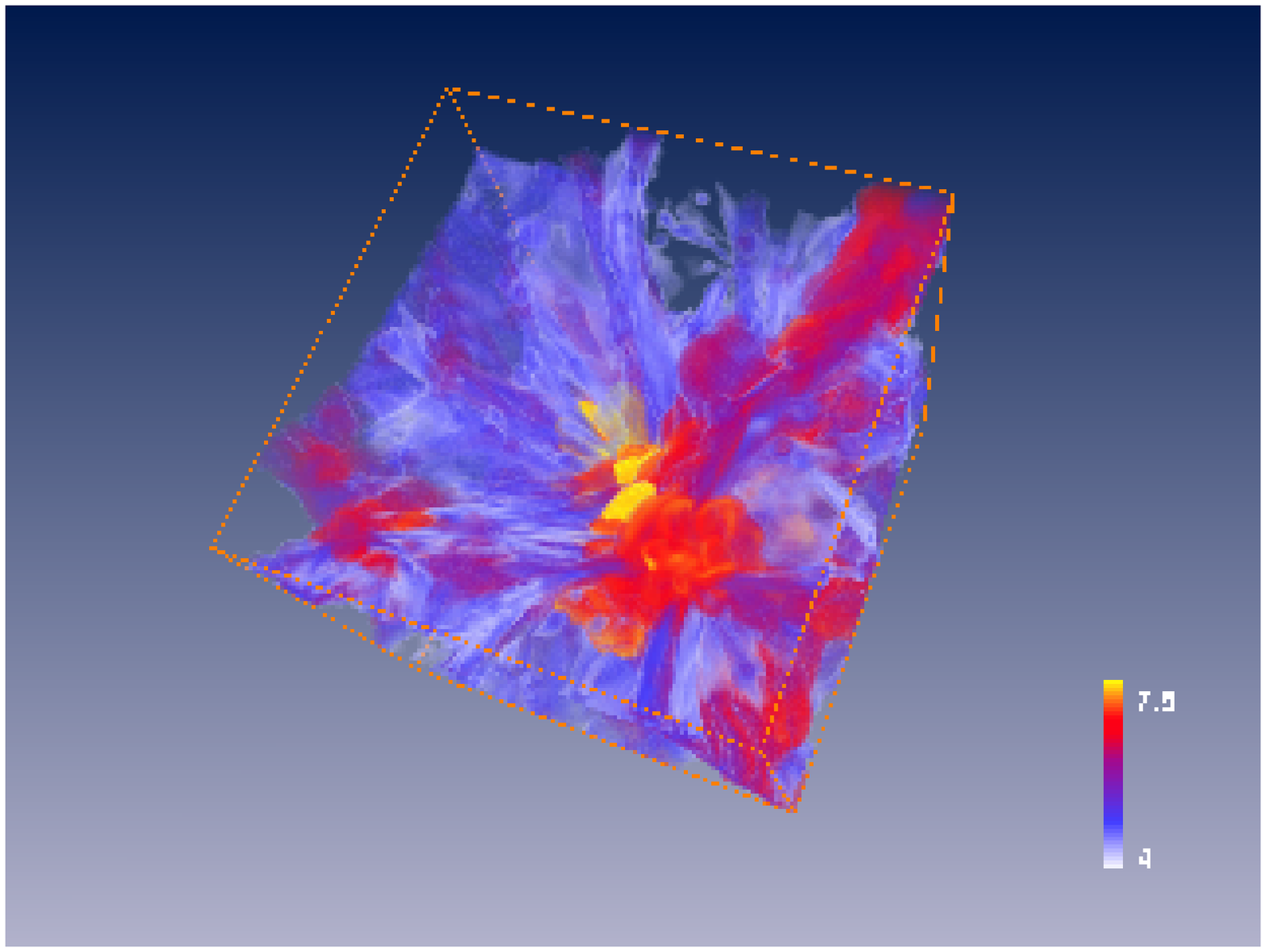}}
\vskip 0.5cm
\caption{Three-dimensional volume rendering of gas temperature in the
$\Lambda$CDM simulation with $1024^3$ grid cells. Top: The box of
$(100 h^{-1} {\rm Mpc})^3$. Middle and Bottom: Two different perspective
views of a $(25 h^{-1} {\rm Mpc})^3$ portion which encompasses the slice
in Fig. 1.}
\end{figure*}

The properties of shock waves associated with cosmic structures have been
studied through a high-resolu\-tion cosmology simulation. For the simulation,
an Eulerian hydro+N-body cosmology code (Ryu et al. 1993) has been used.
The cold dark matter cosmology with a cosmological constant ($\Lambda$CDM)
has been employed with the following parameters:
$\Omega_{BM}=0.043$, $\Omega_{DM}=0.227$, and $\Omega_{\Lambda}=0.73$
($\Omega_{BM} + \Omega_{DM} + \Omega_{\Lambda}=1$),
$h \equiv H_0$/(100 km/s/Mpc) = 0.7, and $\sigma_8 = 0.8$. A cubic region
of size $100 h^{-1}$ Mpc at the current epoch has been simulated inside
a computational box with $1024^3$ gas and gravity cells and
$512^3$ dark matter particles, allowing a spatial resolution of
$97.7 h^{-1}$ kpc. The simulation is the largest of this kind to date.

Fig. 1 illustrates typical cosmic structures found in the simulation,
showing bremsstrahlung X-ray emissivity (upper-left), gas density
(upper-right), gas temperature (lower-left) and shock Mach number
(lower-right) distributions. It represents a slice of
$(25 h^{-1} {\rm Mpc})^2$ centered on a hot cluster in the
simulation box. 
While X-ray clusters are scattered, 
gas density and temperature reveal connected structures
mostly through pancakes and filaments. 
The outer bounds of shock waves follow closely those of gas temperature 
distribution. 
But Mach number distribution shows rich, complex network of weak shock waves 
inside the filaments bounded by strong accretion shocks.
Note that among the protruding structures around the central cluster,
thin features in $T$ and $M$ distribution with a thickness of 
$\sim 1 h^{-1}$ Mpc correspond to cross sections of pancakes, 
while somewhat thick features with a thickness of $\sim 3-5 h^{-1}$ Mpc 
belongs to filaments. 

Fig. 2 shows three-dimensional perspective of cosmic structures through 
volume rendering of gas temperature. Top panel covers the whole simulation
box, while lower two panels represent two different perspectives for
a $(25 h^{-1} {\rm Mpc})^3$ portion centered on the same cluster shown 
in Fig. 1. Images show clearly three topological features, i.e., pancakes,
filaments and knots (clusters) in cosmic structures. Filaments lie
on pancakes, and clusters are distributed along filaments or positioned
at the intersections of filaments. 

Cosmic shocks are most commonly characterized as either accretion shocks,
if they result from infall of diffuse, intergalactic gas onto the
perimeter of clusters, or merger shocks, if they result from collisions
of two clusters. A quick glance at Figs. 1 and 2 reveal that this is an
overly simplified picture. The hierarchical clustering process for structure
formation  produces extremely complex shock structures inside, around
and outside clusters. These shock waves are neither spherical
nor identifiable by simple surfaces. Indeed, they intersect each other,
forming nested shock surfaces, extending from outer bounds of pancakes
and penetrating deep inside clusters. In addition, collisions between
flows in filaments can lead to shocks, and accretion shocks are often
hard to be distinguished from merger shocks, given the complexity that 
accompanies the accumulation of mass in regions where clusters are forming.
Dissipation at these shocks provides the basic heating of the intracluster
medium, although other processes, including feedback from active galactic
nuclei and galaxies may also be important contributors.

According to Miniati et al. (2000) and our analysis of the current simulation,
cosmic shocks come in wide ranges of scales and strengths.
Shock waves have been identified inside, around and outside clusters
on scales from $\la 100$ kpc to a few Mpc. Shock Mach number depends
on the gas temperature, since flow velocities tend generally to be of order
$10^3$ km s$^{-1}$ both inside and outside clusters. Since shocks are found
more frequently inside clusters, where gas temperature is high
($T \ga 10^6$ K), there are more shocks with smaller Much number.
The shock surface-area distribution shows a peak at $M=1$. However, to the
extent that the shocked gas has been virialized, moderate strength
shocks with Mach number roughly in the range $2 \la M \la 3$ would
contribute most to heating of the intracluster medium. 

Cosmic shocks are capable of accelerating cosmic rays efficiently, so
deserve a close scrutiny in that regard.  It is important to remember that,
since cosmic rays are effectively tied to the intracluster medium up
to pretty high energies, the integrated shock history of the intracluster
medium determines the character of the cosmic ray population.
So, one would not expect cosmic rays produced in this way to be only
associated with recent merger events, for example. Our analysis of
the simulation indicates the shocks, which contribute most to
acceleration of cosmic rays, would be those in the range $3 \la M \la 4$.
Miniati et al. (2001) and our analysis show such shock waves can accelerate
cosmic ray protons with pressure up to $\sim50\%$ of gas thermal pressure.

Details of the simulation in this section will be reported elsewhere
(Ryu et al. 2002b; Kang et al. 2002).

\section{COSMOLOGY SIMULATIONS WITH\\ COSMIC RAY DYNAMICS}

The simulations used to study the dynamical effects of cosmic rays
have also employed the $\Lambda$CDM cosmology with the parameters same as those
in the previous section, and have been performed with the same cosmology
code. However, a cubic region of smaller comoving size,
$75h^{-1}$ Mpc, has been chosen for computational domain, and smaller numbers
of cells and particles, $512^3$ cells for gas and gravity and $256^3$
particles for dark matter, have been used. Two simulations have been done,
one with cosmic ray dynamics included and the other without cosmic rays.

The following equation for cosmic ray pressure has been solved
in addition to the usual standard set of equations for dark matter 
and gas (Ryu et al. 1993),
\begin{equation}
{\partial p_{CR} \over \partial t} + {1 \over a} u_k {\partial P_{CR}
\over \partial x_k} + {\gamma_{CR} \over a} P_{CR} {\partial u_k \over
\partial x_k} = - 3(\gamma_{CR}-1){\dot a \over a} p_{CR}.
\end{equation}
Here, $a$ is the expansion parameter, $x_k$ is the comoving length,
$p_{CR}$ is the comoving pressure of cosmic rays, $u_k$ is the proper
peculiar velocity, and $\gamma_{CR}$ is the adiabatic index of 
cosmic rays which has been assumed to be 4/3. Diffusion of cosmic rays
has been ignored, since the diffusion length is much shorter than the
computational cell size for most of cosmic rays. The dynamical feedback
of cosmic rays to gas has been incorporated by including the
$(1/a)(\partial P_{CR}/ \partial x_k)$ term in the momentum equation,
along with the $(1/a)u_k(\partial P_{CR}/\partial x_k)$ term in the
energy equation.

It has been assumed that the sources, which deposit cosmic ray energy
into the intracluster medium, form at 40 different epochs after
the redshift $z=10$, if the following criteria
are satisfied in each grid cell
\begin{equation}
M_{gas} \geq {3\times10^{10}\over1+z}h^{-1} M_{\odot},~~~
{\partial u_k \over \partial x_k} < 0,
\end{equation}
where $M_{gas}$ is the total gas mass inside the cell. It has been
further assumed that each source ejects the following amount
of cosmic ray energy into the intracluster medium
\begin{equation}
E_{CR} = 3\times10^4 h^{-1} M_{\odot} \times c^2.
\end{equation}
Note that at $z=0$ this translates into the cosmic ray efficiency,
$E_{CR}/ M_{gas}c^2 = 10^{-6}$.

Bottom panel of Fig. 3 shows that the source formation history, which has
been realized by the criteria (2), increases first, and
then stays more or less constant after $z\sim2$. Total of $\sim4\times10^4$
sources have formed in the simulation box of $(75h^{-1}{\rm Mpc})^3$. 
Middle panel shows the resulting mass averaged densities of gas
kinetic energy, gas thermal energy and cosmic ray energy, which reflect
mostly the averaged energy densities inside clusters with high mass density.
In clusters the cosmic ray energy dominated over the gas thermal energy before
$z\sim 2$ in our model, during which the gas thermal energy was still small.
After $z\sim 2$, the ratio of cosmic ray to gas thermal energies has decreased.
The cosmic ray energy density has become and stays $\sim 50\%$ of the gas
thermal energy density after $z\sim1$. With such amount of cosmic ray energy,
the gas density perturbation has reduced by $\sim 20\%$ at the present
epoch, while the density perturbation of dark matter has been hardly
affected, as shown in top panel.

Fig. 4 shows the power of density perturbation at three different epochs.
It has been decreased by $\sim 25\%$ at $z=2$, by $\sim 35\%$ at $z=1$,
and by $\sim 45\%$ at $z=0$ on the cluster scale of $\sim 1 h^{-1}$ Mpc,
due to the dynamical effects of cosmic ray pressure. However, the
structures of scales larger than the cluster scale have been less
disturbed, since sources form mostly at the highest density peaks
within clusters.

To study the dynamical effects of cosmic ray pressure on individual
clusters, we have randomly identified several clusters in the computational
box, and compared their properties in the cases with/without cosmic ray
dynamics. Cosmic ray pressure provides additional support against
infalling flows. As a result, the intracluster medium becomes less
concentrated with cosmic rays, and the virial temperature reduces to
lower values. The clusters with $E_{CR} \sim 1/2 E_{th}$ have been
significantly modified. Their X-ray bremsstrahlung emission and
temperature decreased by $\sim 50\%$.

Further details of the simulations in this section will be reported 
in Ryu et al. (2002a).

\begin {figure*}
\vskip 0cm
\centerline{\epsfysize=20cm\epsfbox{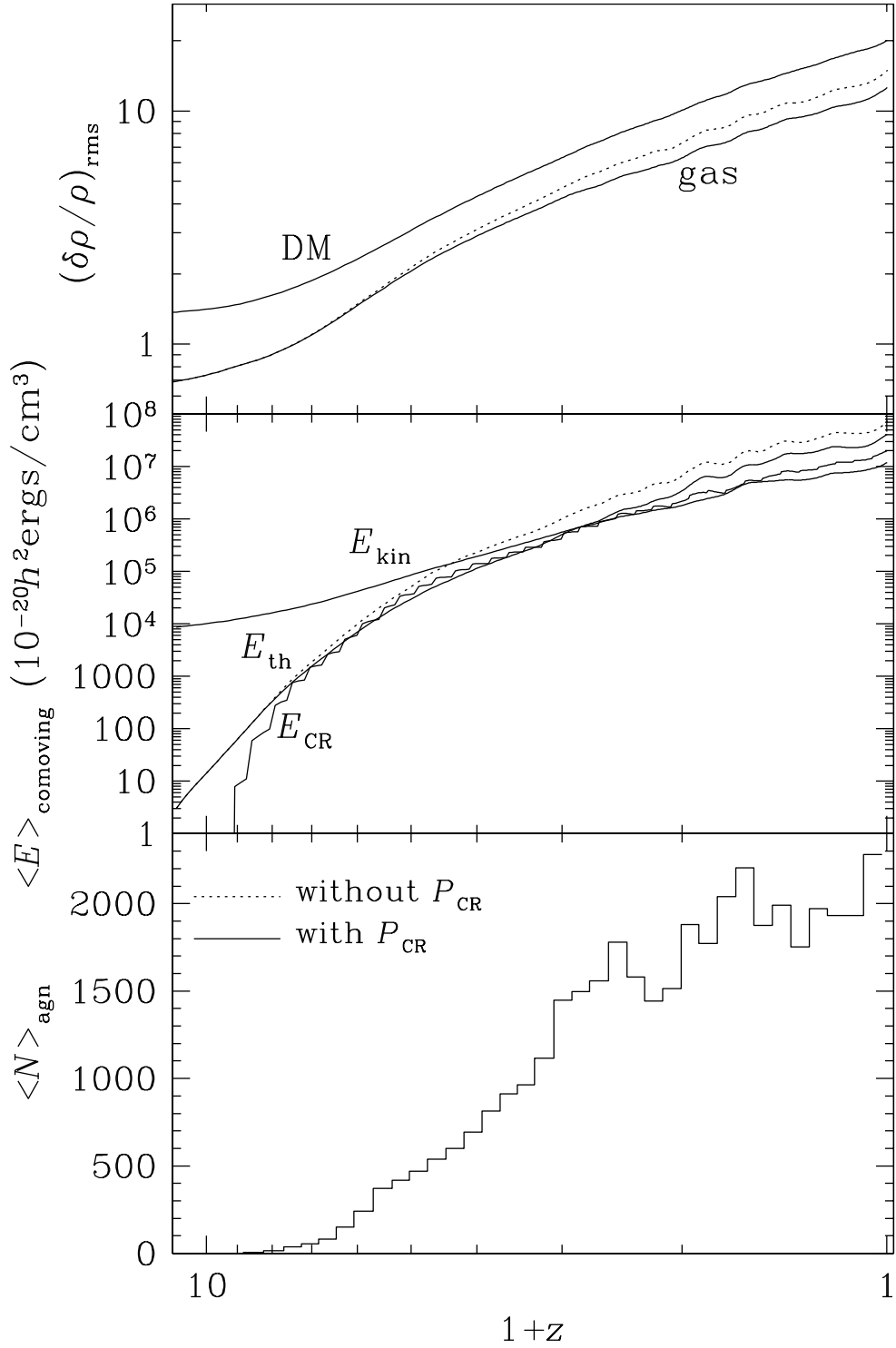}}
\vskip 0cm
\caption{Top: Temporal evolution of the root-mean-square density
perturbations of dark matter and gas in the $512^3$ $\Lambda$CDM
simulations with (solid line) or without (dotted line) cosmic ray energy
ejection from AGN-like sources. Middle: Temporal
evolution of the mass-averaged energy per unit comoving volume.
$E_{kin}$ is the gas kinetic energy, $E_{th}$ is the gas thermal
energy, and $E_{CR}$ is the cosmic ray energy. Bottom: Formation
history of cosmic ray sources. Sources are assumed to form at 40
logarithmically spaced epochs after $z=10$.}
\end{figure*}

\begin {figure*}
\vskip 0cm
\centerline{\epsfysize=20cm\epsfbox{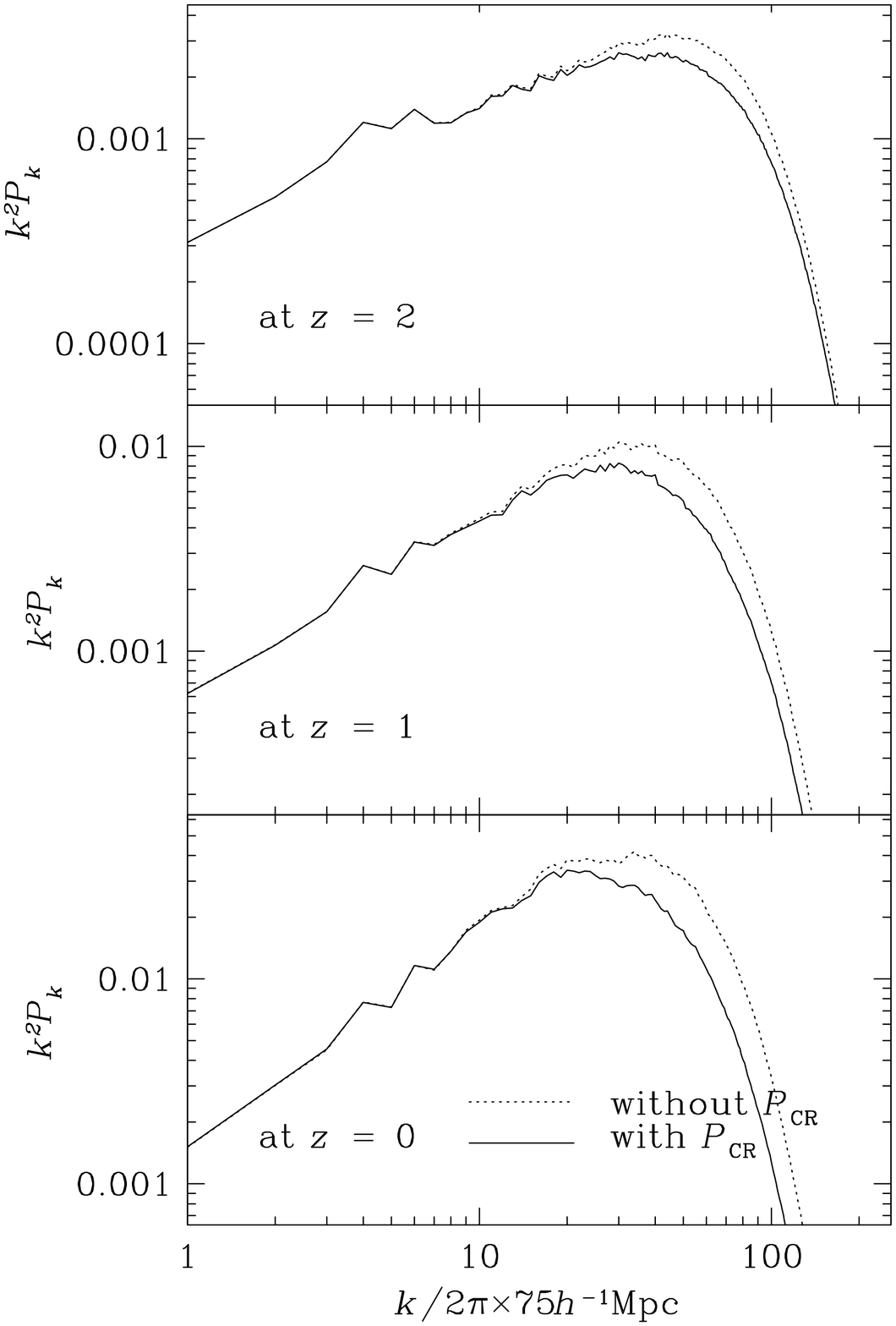}}
\vskip 0cm
\caption{Power spectrum of gas density perturbations at three different
epochs in the $512^3$ $\Lambda$CDM simulations with (solid line) or
without (dotted line) cosmic ray energy ejection from AGN-like sources.} 
\end{figure*}

\vskip 0.4cm
This work was supported by grant No. R01-1999-000-00023-0
from the Korea Science \& Engineering Foundation.
Simulations have been made through a support by
the Grand Challenge Program of KISTI Supercomputing Center.

\end{document}